# Scintillation and irregularities from the nightside part of a Sun-aligned polar cap arc

Christer van der Meeren[1], Kjellmar Oksavik[1,2], Dag A. Lorentzen[3,4], Larry J. Paxton[5], and Lasse B. N. Clausen[6]

[1]Birkeland Centre for Space Science, Department of Physics and Technology, University of Bergen, Bergen, Norway, [2]University Centre in Svalbard, Longyearbyen, Norway, [3]Birkeland Centre for Space Science, University Centre in Svalbard, Longyearbyen, Norway, [4]Now at British Antarctic Survey, Cambridge, UK, [5]The Johns Hopkins University Applied Physics Laboratory, Laurel, Maryland, USA, [6]Department of Physics, University of Oslo, Oslo, Norway

**Abstract** In this paper we study the presence of irregularities and scintillation in relation to the nightside part of a long-lived, Sun-aligned transpolar arc on 15 January 2015. The arc was observed in DMSP UV and particle data and lasted at least 3 h between 1700 and 2000 UT. The arc was more intense than the main oval during this time. From all-sky imagers on Svalbard we were able to study the evolution of the arc, which drifted slowly westward toward the dusk cell. The intensity of the arc as observed from ground was 10–17 kR in 557.7 nm and 2–3.5 kR in 630.0 nm, i.e., significant emissions in both green and red emission lines. We have used high-resolution raw data from global navigation satellite systems (GNSS) receivers and backscatter from Super Dual Auroral Radar Network (SuperDARN) radars to study irregularities and scintillation in relation to the polar cap arc. Even though the literature has suggested that polar cap arcs are potential sources for irregularities, our results indicate only very weak irregularities. This may be due to the background density in the northward IMF polar cap being too low for significant irregularities to be created.

## 1. Introduction

The high-latitude ionosphere is highly dynamic, forming plasma irregularities on a wide variety of scale sizes, from 1000 km scale islands of enhanced plasma density down to irregularities at decameter scale [e.g., *Weber et al.*, 1984; *Basu et al.*, 1990a, 1990b; *Oksavik et al.*, 2012; *Moen et al.*, 2012, 2013].

Southward interplanetary magnetic field (IMF) conditions are thoroughly investigated and are the focus of most irregularity studies. During southward IMF conditions one can frequently observe polar cap patches, tongues of ionization, poleward moving auroral forms, and magnetospheric substorms. Polar cap patches are 100–1000 km islands of enhanced plasma density being segmented from the dayside high-density plasma in the cusp region [e.g., *Weber et al.*, 1984; *Lockwood and Carlson*, 1992; *Oksavik et al.*, 2010; *Carlson*, 2012; *Zhang et al.*, 2013a, 2013b]. Patches may develop smaller-scale irregularities down to decameter scale through the Kelvin-Helmholtz (KH) and gradient drift instabilities [e.g., *Basu et al.*, 1990a; *Carlson et al.*, 2007, 2008; *Moen et al.*, 2012; *Oksavik et al.*, 2012; *Clausen et al.*, 2016]. During strong and stable polar cap convection, segmentation may not happen and a continuous tongue of ionization (TOI) may be formed across the polar cap [*Sato*, 1959; *Knudsen*, 1974; *Foster et al.*, 2005]. In addition to patches, transient magnetopause reconnection gives rise to poleward moving auroral forms (PMAFs) [e.g., *Feldstein and Starkov*, 1967; *Vorobjev et al.*, 1975; *Sandholt et al.*, 1990, 1993; *Thorolfsson et al.*, 2000]. Magnetospheric substorms cause an explosive release of energy into the auroral ionosphere [e.g., *Akasofu*, 1964; *McPherron*, 1970, 1979; *Rostoker et al.*, 1980; *Elphinstone et al.*, 1996]. All of these phenomena are associated with irregularities causing disturbances in transionospheric radio links: Patches [*Buchau et al.*, 1985; *Weber et al.*, 1986; *Basu et al.*, 1990a, 1991, 1994, 1998; *Coker et al.*, 2004; *Clausen et al.*, 2016], TOIs [*van der Meeren et al.*, 2014], PMAFs [*Oksavik et al.*, 2015; *Jin et al.*, 2015], and auroral activity [*Aarons et al.*, 2000; *Spogli et al.*, 2009; *Prikryl et al.*, 2010; *Ngwira et al.*, 2010; *Tiwari et al.*, 2012; *Jiao et al.*, 2013; *Kinrade et al.*, 2013; *Hosokawa et al.*, 2014; *van der Meeren et al.*, 2015; *Clausen et al.*, 2016].

However, northward IMF conditions are much less studied, although they occur half the time. During such conditions, the predominant feature of the polar ionosphere is polar cap arcs [e.g., *Carlson*, 1994, and references therein]. While reviewing space weather challenges for the polar cap ionosphere, *Moen et al.* [2013] suggested in passing that flow shears near transpolar arcs might be a space weather concern by creating irregularities through the KH instability. Unfortunately, there exists barely any studies at all on irregularities from polar cap arcs. In this paper, following the general methods of *van der Meeren et al.* [2014, 2015] and *Oksavik et al.* [2015], we use all-sky images, raw and reduced data from several multiconstellation global navigation satellite system (GNSS) receivers, and Super Dual Auroral Radar Network (SuperDARN) backscatter, as well as DMSP UV and particle data to study scintillation and irregularities in relation to an intense, detached polar cap arc.

First, we will give a brief overview of polar cap arcs, GNSS scintillations, and the current state of knowledge of the relation between these two phenomena.

### 1.1. Polar Cap Arcs

Optical arcs poleward of the auroral oval have been observed for at least a century [*Mawson*, 1916] and primarily occur during quiet geomagnetic conditions [*Davis*, 1963] and northward IMF [*Berkey et al.*, 1976; *Gussenhoven*, 1982]. The phenomenon is not yet fully understood.

The naming is not entirely consistent in literature. Usually [e.g., *Fear and Milan*, 2012a, 2012b; *Carlson*, 1994; *Newell et al.*, 2009, and references therein], the term "polar cap arcs" and "high-latitude arcs" refer to any type of aurora inside the polar cap. Such auroral features are often Sun-aligned and may stretch across a significant part of the polar cap, giving rise to the names Sun-aligned arcs and transpolar arcs, respectively. In the case that they connect the nightside and dayside auroral oval, they are often termed theta aurora. In this paper we will use the term "polar cap arc."

According to a review by *Newell et al.* [2009], three different types of polar cap arcs exist, possibly with different underlying mechanisms. The first type is intensifications of polar rain. These are common, but weak, and consist of only electron precipitation (without associated ions). The second type is Sun-aligned arcs which appear detached from the auroral oval in optical data but are adjacent to the auroral oval in particle data (though usually with a plasma regime distinct from the auroral oval). The third type occurs very rarely, and is intense, Sun-aligned arcs well detached from the auroral oval in both particle and optical data. These events can include plasma sheet ions (such as $O^+$) as well as electrons.

Convection has been observed to be antisunward in the nightside portion of the arc and either sunward [*Eriksson et al.*, 2006] or mixed [*Liou et al.*, 2005] in the dayside portion of the arc. In the vicinity of polar cap arcs, convection is structured, with multiple reversals, flow shears, and flow channels [e.g., *Carlson and Cowley*, 2005; *Eriksson et al.*, 2006; *Zou et al.*, 2015].

### 1.2. Scintillation From Polar Cap Arcs

Scintillations are rapid fluctuations in the amplitude or phase of a radio signal, such as GNSS signals, and can be caused by irregularities in the ionosphere with scale sizes of decameters to kilometers [e.g., *Hey et al.*, 1946; *Basu et al.*, 1990a, 1998; *Kintner et al.*, 2007]. The irregularity scale sizes causing scintillation are determined by the Fresnel radius, which for GPS L1 signals (1575.42 MHz) passing through irregularities at an altitude of 350 km is approximately 360 m [*Forte and Radicella*, 2002]. Amplitude scintillations are represented by the $S_4$ index and are due to irregularities with scale sizes at and below the Fresnel radius, from hundreds of meters down to tens of meters. Phase scintillations are represented by the $\sigma_\phi$ index and are caused by irregularities above the Fresnel radius, from a few hundred meters to several kilometers [e.g., *Kintner et al.*, 2007].

*Buchau et al.* [1985] and *Basu et al.* [1990b] observed 250 MHz scintillation in relation to polar cap arcs. The latter study found an increase in $\sigma_\phi$ of ∼7 times the background level and attributed it to sheared plasma flow in association with the arc. Scintillation from polar cap arcs at GNSS *L* band frequencies (∼1.2–1.6 GHz) is less studied. *Prikryl et al.* [2015a] included one example of moderate to strong GPS phase scintillation from the dayside part of a transpolar arc following a strong solar wind dynamic pressure pulse which resulted in large $B_y$ oscillations. *Prikryl et al.* [2015b] performed a statistical study of high-latitude scintillation and found that for northward IMF, the $B_y$-dependent dayside dawn-dusk asymmetry in scintillation was consistent with the expected asymmetry of Sun-aligned arcs. The statistical scintillation in the duskside and dawnside sectors was then attributed to Sun-aligned arcs.

We have not been able to find any studies of GNSS scintillation from polar cap arcs in the nightside ionosphere.

## 2. Instrumentation
### 2.1. GNSS Receivers

The GNSS data used in this study come from two NovAtel GPStation-6 GNSS Ionospheric Scintillation and TEC Monitor. The receivers were installed in Svalbard in 2013 and are operated by the University of Bergen.

The locations, geographic latitudes (GLAT), geographic longitudes (GLON), and magnetic latitudes (MLAT) of the receivers are Longyearbyen (78.1° GLAT, 16.0° GLON, and 75.4° MLAT) and Ny-Ålesund (78.9° GLAT, 11.9° GLON, and 76.4° MLAT). Magnetic midnight is around 2100 UT at these locations.

All receivers track the GPS, GLONASS, and Galileo constellations at several frequencies. GPS and GLONASS satellites were visible during the period of study and were tracked at L1 (1575.42 MHz). Additionally, one GPS satellite was tracked at L5 (1176.45 MHz), and GLONASS was tracked at L2 (1227.60 MHz). The different signals showed no significant difference in scintillation intensity (the levels were marginally higher in the lower frequencies), so L1 will be used in this study.

The receivers output 60 s averaged data. The $\sigma_\phi$ index is computed by detrending the carrier phase using a sixth-order Butterworth filter with a cutoff frequency of 0.1 Hz and computing the standard deviation over periods of 60 s. The $S_4$ index is computed by taking the standard deviation of the received power over periods of 60 s and normalizing by its mean value.

Using a 0.1 Hz cutoff frequency for the $\sigma_\phi$ index is prevalent in the literature [e.g., *Mitchell et al.*, 2005; *Béniguel et al.*, 2009; *Li et al.*, 2010; *Forte et al.*, 2011; *Gwal and Jain*, 2011; *Alfonsi et al.*, 2011; *Garner et al.*, 2011; *Kinrade et al.*, 2012; *Jiao et al.*, 2013; *Jin et al.*, 2014]. The $\sigma_\phi$ index has been shown to be highly sensitive to the detrending filter cutoff frequency, and 0.1 Hz can be problematic in the high-latitude regions [e.g., *Forte*, 2005]. However, as performed by *van der Meeren et al.* [2014, 2015] and *Oksavik et al.* [2015], it is possible to use a spectrogram of the raw phase to get a better overview of the phase variations at different scales. The spectrograms were made using wavelet analysis, based on software provided by *Torrence and Compo* [1998]. In line with previous GNSS studies, the Morlet wavelet was chosen as the mother wavelet [e.g., *Mushini et al.*, 2012]. The wavelet spectrograms require no detrending of the raw phase. For further details on wavelet analysis the reader is referred to the literature [e.g., *Torrence and Compo*, 1998; *Mushini et al.*, 2012].

The receivers also output 50 Hz raw data (phase and power). In this study, the raw data are used to compute the $\sigma_\phi$ index with 1 s resolution using the method described above with periods of 1 s instead of 60 s.

For each receiver, data from satellites were excluded in directions (elevation/azimuth) known to cause multipath problems for that receiver. No satellites were used below 20° elevation. Occasionally, the receivers glitch (most likely due to clock adjustments), causing a transient increase in scintillation for all satellites seen in that receiver. Scintillation data around these glitches have been removed in this study.

The two all-sky imagers used in this study (to be detailed below) are colocated with the GNSS receivers. Therefore, we have not performed any geographical projection of the GNSS ionospheric piercing points. We have instead shown ground-based optical data and associated GNSS data on a polar grid using elevation and azimuth.

### 2.2. All-Sky Imagers

We use data from two all-sky imagers (ASI). The ASIs are colocated with the GNSS receivers in Ny-Ålesund and Longyearbyen and are operated by the University of Oslo. Both imagers are calibrated and have filters for red (630.0 nm) and green (557.7 nm) emissions. Emissions at 630.0 nm are projected to 250 km, while emissions at 557.7 nm are projected to 150 km altitude. The exact projection altitude is not important for how the optical data are used in this study.

For most of the time under study, parts of the 557.7 nm data were of relatively poor quality due to technical problems (though the arc was clearly visible throughout the interval). We have, therefore, chosen to show mostly 630.0 nm images. These problems did not influence the line-of-sight data shown in Figures 5 and 6.

### 2.3. DMSP UV and Particle Data

The Defense Meteorological Satellite Program (DMSP) spacecraft are in polar, Sun-synchronous orbits at a nominal altitude of 840 km. The Special Sensor Ultraviolet Spectrographic Imager (SSUSI) instrument observes UV emissions at several wavelengths [*Paxton et al.*, 1992]. The channel used in the current study is 140–150 nm (Lyman-Birge-Hopfield short filter, LBHS). We use calibrated, background-corrected intensities from the Sensor Data Record (SDR) product. The data are projected to 150 km altitude. The projection is only used to show that the satellite- and ground-based optical data are consistent, and the exact choice of altitude is not critical for the purposes of this study.

The DMSP SSJ/4 particle detectors always point toward the zenith and measure precipitating electrons and ions between 30 eV and 30 keV.

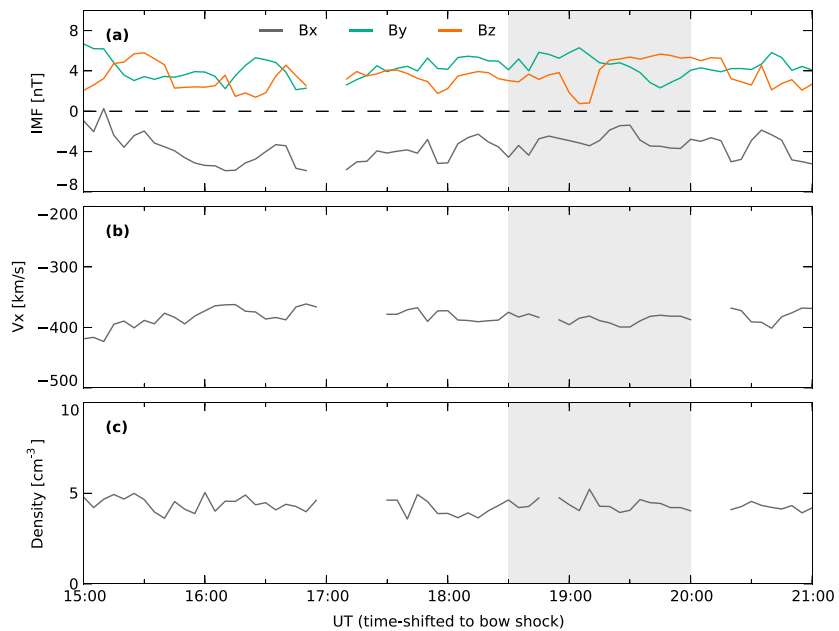

**Figure 1.** Solar wind conditions on 15 January 2015 several hours prior to, during, and after the event. The data show (a) northward IMF, (b) stable speed, and (c) stable density. The approximate time under study is shaded in gray.

### 2.4. Solar Wind

We use plasma and IMF data from the Wind spacecraft [*Lepping et al.*, 1995; *Ogilvie et al.*, 1995]. The spacecraft was located at $(X, Y, Z) = (196, 37, -15)$ $R_E$ (geocentric solar ecliptic coordinates). The data are time shifted to the bow shock by the NASA OMNIWeb service.

### 2.5. SuperDARN Data

The Super Dual Auroral Radar Network (SuperDARN) transmits signals in the high-frequency (HF) range, which scatters coherently from field-aligned decameter-scale irregularities [*Greenwald et al.*, 1995; *Chisham et al.*, 2007]. HF backscatter is, therefore, an indicator of the presence of irregularities with scale sizes of tens of meters.

### 3. Observations

Figure 1 shows the solar wind conditions around the time of the event on 15 January 2015. The time under study is shaded in gray. The IMF is northward with $B_y$ positive several hours prior to the event. The solar wind conditions, in general, are stable. Figure 2 shows a polar cap arc in both optical and particle data (the particle data are from the dayside part of the arc; unfortunately, there were no DMSP passes of the nightside part of the arc). The arc is very long-lived, lasting at least 3 h from 1700 to 2000 UT. In the spectrograms, diffuse aurora is visible at the equatorward edges, with discrete auroral structures farther poleward followed by polar rain. The arc precipitation is outlined with black borders. The arc exceeds the main oval in optical intensity (LBHS) and shows similar fluxes and higher electron energies than the discrete part of the main oval. The inverted V structures in the arc (most clear in Figure 2d but visible in Figures 2e and 2f as well) indicate field-aligned acceleration of electrons with resulting energies of up to ∼3 keV. High-energy ions (∼10 keV) are also visible in the arc, though the ion flux is very low. The dayside part of the arc seems to be broken up into multiple arcs in both the optical and particle data. In the nightside optical data (supported by the ground-based optical data to be shown later), the arc seems not to be broken up.

The stability and intensity of the arc makes this an ideal candidate for a detailed case study. It is also interesting to note that the arc seems well separated from the main oval in both optical and particle data, and this may thus be the rare "type 3" arc as described by *Newell et al.* [2009].

Figures 3a and 3b show a good match between ground-based (from Longyearbyen) and satellite-based optical data. The arc is clearly visible in both red (630.0 nm) and green (557.7 nm) emissions throughout the interval under study. Figures 3c–3e show how the arc moves across the ASI field of view in elevation/azimuth coordinates. The movement across the ASI field of view is partly because of the rotation of the Earth and partly because the arc drifts toward dusk. The movement is irregular: It seems stationary in a magnetic local time (MLT) frame until around 1905 UT. After this it drifts westward at a speed of around 600–900 m/s until it starts to fade around 2000 UT, but with intermittent stops and slight bulges along the way which makes it difficult to pin down an exact drift speed. The correction for the rotation of the Earth is around 100 m/s and is not important in this context. The intensity of the arc varies during the event. The maximum intensity as measured by the ASI is 10–17 kR in 557.7 nm and 2–3.5 kR in 630.0 nm. These are significant emission intensities, and one should expect some scintillation [*Kinrade et al.*, 2013].

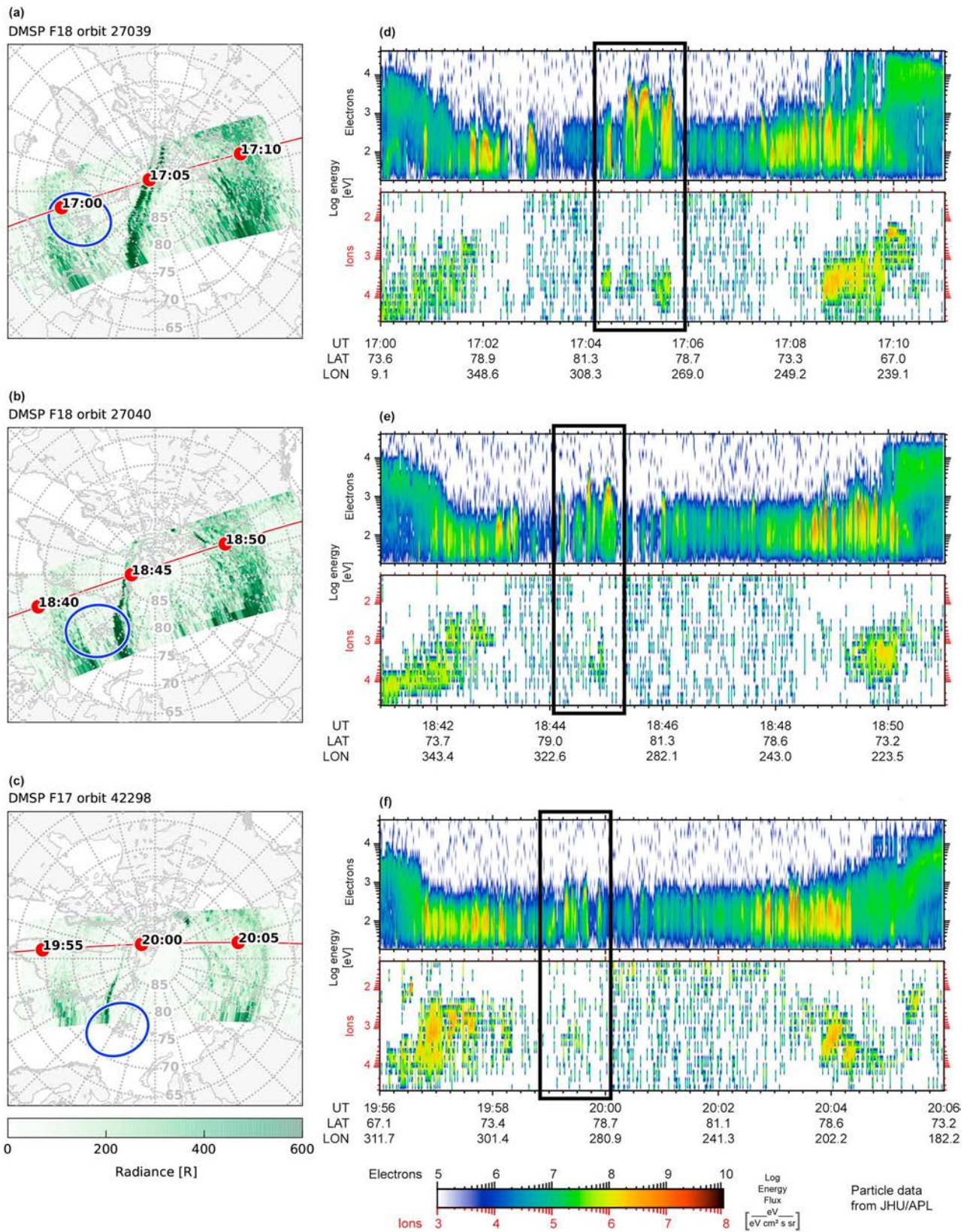

**Figure 2.** (a–c) Optical data and (d–f) particle spectra from three passes of DMSP satellites on 15 January 2015. The data show a stable transpolar Sun-aligned arc lasting for at least 3 h, well separated from the auroral oval in both optical and particle data. The field of view of the Longyearbyen ASI at 250 km altitude is shown as a blue circle. The optical data are projected onto a geomagnetic grid. Black boxes outline the (dayside) polar cap arc in the particle data.

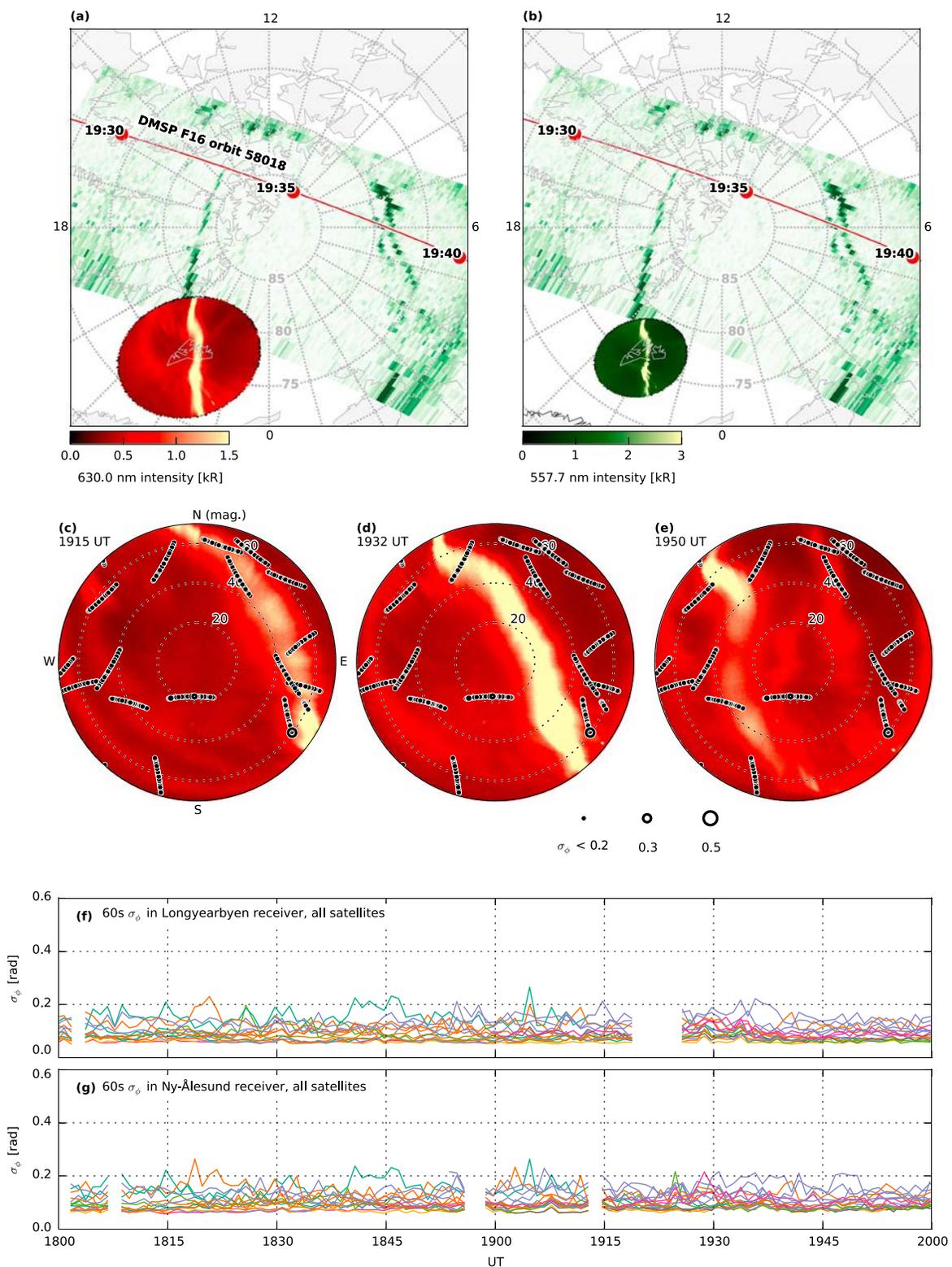

**Figure 3.** (a–b) Ground- and space-based optical data from the event on 15 January 2015. The arc is visible from the ground in both 630.0 nm and 557.7 nm data. (c–e) The movement of the arc across the all-sky field of view, displayed on an azimuth/zenith angle grid. Magnetic north is up, magnetic west is left. Scintillation data from all visible satellites between 1912 and 1952 UT are shown as traces of dots/circles (identical for all three images). The circles below the images indicate the scintillation scale. (f–g) Scintillation data from all visible satellites in Longyearbyen and Ny-Ålesund. Even though the arc crosses most satellites in both receivers, scintillation is very low at all times.

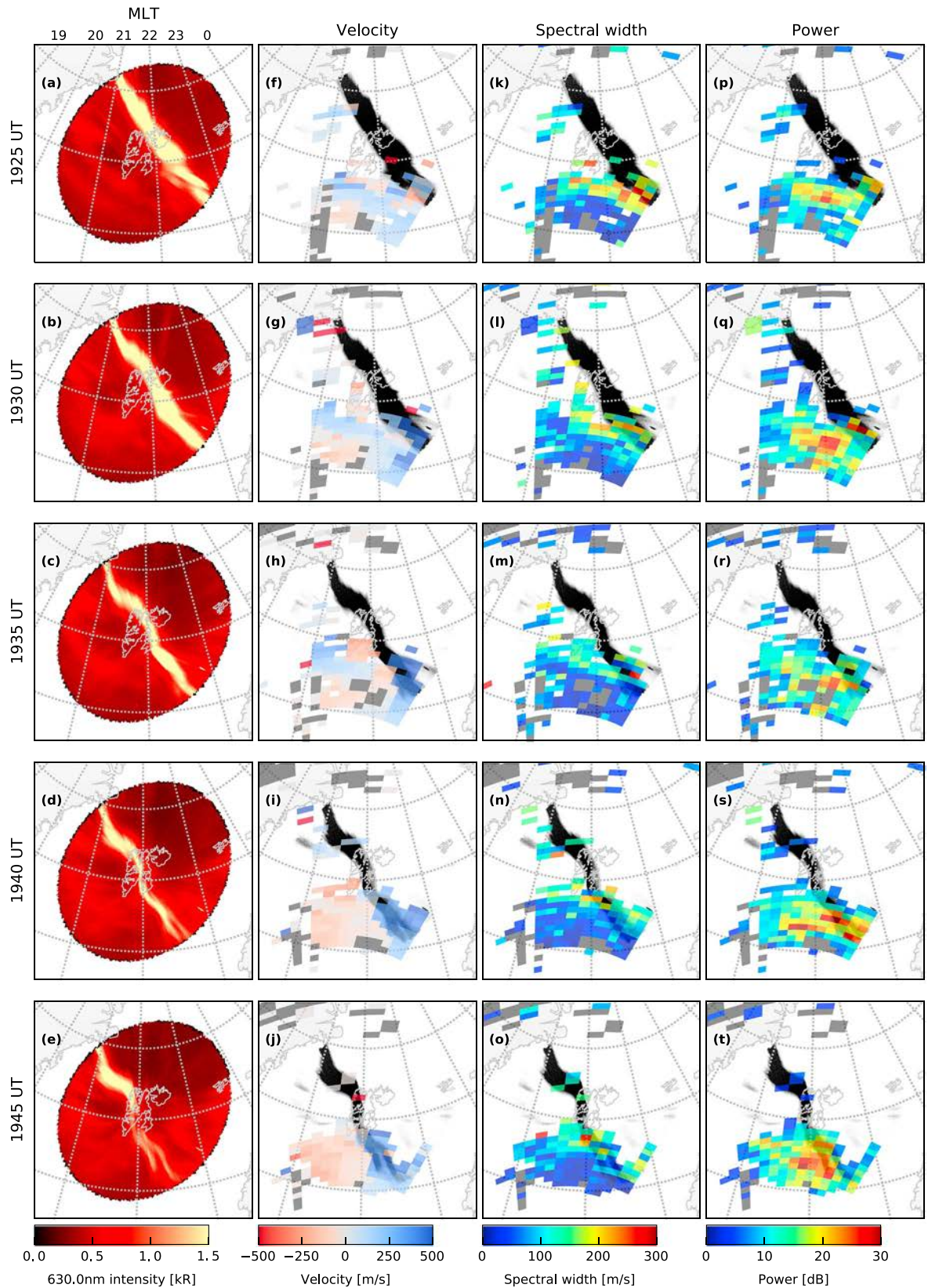

**Figure 4.** Backscatter data from the Hankasalmi SuperDARN radar at five different times during the event on 15 January 2015. (a–e) Optical data for reference. The arc is shown in black in the other panels. (f–j) Drift velocity. (k–o) Spectral width. (p–t) Backscatter power. Observe the flow shear in the velocity data (Figures 4i–4j), as well as the enhanced spectral width and backscatter power near the arc.

Overlaid on top of Figures 3c–3e are tracks of phase scintillation for all the visible satellites in the Longyearbyen receiver between 1912 and 1952 UT (identical for all three panels). This shows that phase scintillation is almost always below $\sigma_\phi = 0.2$. Figures 3f and 3g show phase scintillation for all satellites in both Longyearbyen and Ny-Ålesund. Phase scintillation is almost always below 0.2, only barely exceeding 0.2 for short amounts of time. This shows clearly that there is no or only very weak scintillation from the arc. A more detailed look at scintillation in relation to line-of-sight intensities for specific satellites will be given shortly.

Figure 4 shows backscatter data from the Hankasalmi SuperDARN radar. The auroral images (Figures 4a–4e) are shown for reference, and the arc itself is shown in black in the other panels. The presence of backscatter near the arc indicates field-aligned irregularities at decameter scale. There is also enhanced spectral width (∼200 m/s) and backscatter power (∼25–30 dB) in relation to the arc. The velocity data (Figures 4f–4j, especially Figures 4i–4j) show a clear velocity shear in relation to the arc. At 22 MLT (where the radar observations are made) during strongly positive IMF $B_y$, one would expect steady westward flow which would result in flows toward the radar in the east and away from the radar in the west [e.g., *Ruohoniemi et al.*, 1989]. However, the flow shear we observe moves with the arc and seems sufficiently abrupt (350 m/s toward versus 50 m/s away in two adjacent range gates 350 km apart) that this is likely a proper flow shear associated with the arc.

Next we will provide a more detailed look at scintillation data for six selected satellites. Figure 5 shows the first three, which are GPS 11, GLONASS 20, and GPS 28. Figures 5a–5c and 5g–5i show emission intensities in the line of sight to the satellites from the Longyearbyen and Ny-Ålesund ASIs, respectively. The intensities are based on the median of a 7-by-7 pixel window centered on the satellite. The arc moves slowly at first, which is why GPS 11 spends a long time in the emissions. The arc passes GLONASS 20 and GPS 28 much more quickly, both between 1920 and 1930 UT. The maximum emission intensities in the line of sight to the satellites vary between 4 and 16 kR for 557.7 nm and are consistently around 2 kR for 630.0 nm.

Figures 5d–5f and 5j–5l show wavelet power spectra of 50 Hz raw phase. In the satellite/receiver combinations where the passing of the arc is clearest and most intense, namely, Figures 5e, 5f, and 5l, slight enhancements of phase variations are visible over the background spectrum, extending down to 1 s variations in Figures 5f and 5l. These enhancements are not particularly severe, and the scintillation data reflect this: Figures 5m–5o show amplitude scintillation, which stays very low throughout the passing of the arc, normally around 0.1 and at all times below 0.2. Figures 5p–5r show phase scintillation, which during the crossing of the arc is slightly enhanced in GLONASS 20 and GPS 28 (Figures 5q–5r) but only up to $\sigma_\phi = 0.2$ which is still only very weak scintillation.

Figures 5s–5u show vertical total electron content (VTEC) from the three satellites. A clear enhancement is visible during the passing of the arc. It is most clear in GPS 28 (Figure 5u), where a 5 total electron content unit (TECU; 1 TECU = $10^{16}$ el m$^{-2}$) increase can be seen. This is effectively an increase in TEC by a factor of 2. It is unknown what the earlier VTEC enhancement in Ny-Ålesund is for this satellite.

Figure 6 tells much the same story for three different satellites at a later time. The passing of the arc is clearly visible in the optical line-of-sight intensities (Figures 6a–6c and 6g–6i), though very weak in GPS 24. When crossing GPS 24 the arc had faded and the intensity varied greatly along the arc (as shown in Figure 3e), with a dim part of the arc crossing the satellite. The spectrograms (Figures 6d–6f and 6j–6l) show no or only very slight enhancements during the arc crossing. Amplitude scintillation stays around $S_4 = 0.1$ at all times, and phase scintillation stays below $\sigma_\phi = 0.2$, effectively meaning there is no or only very weak scintillation at all (though some slight enhancements over the background level can still be seen at times). The VTEC data again show enhancements of 2–5 TECU in relation to the arc.

We would like to end this section by pointing out that other satellites besides these six were studied in the same manner and support these findings. Additionally, we have briefly looked at four other polar cap arc

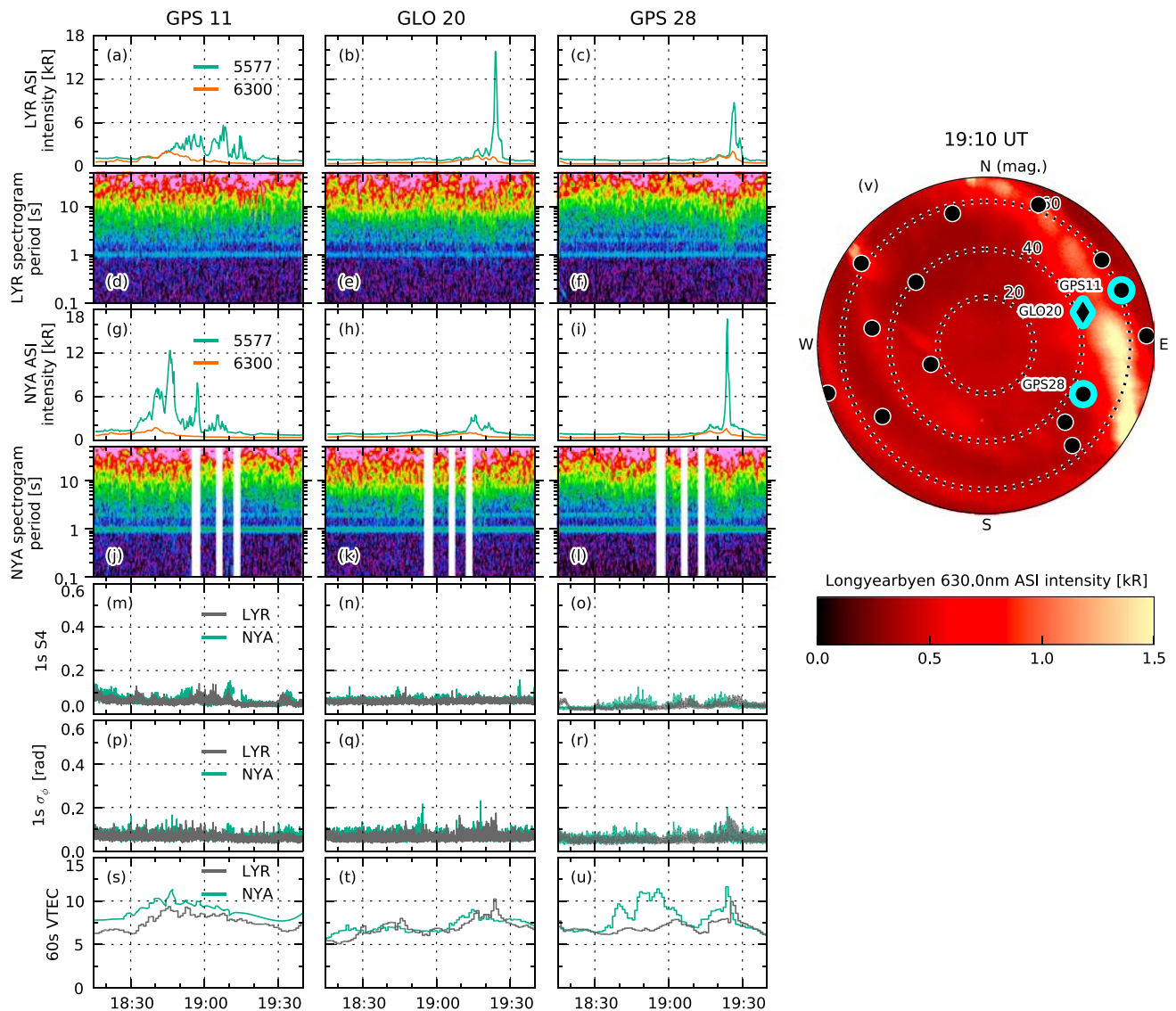

**Figure 5.** GNSS scintillations, vertical total electron content (VTEC), and phase variations for three selected satellites in relation to line-of-sight auroral intensity during the event on 15 January 2015. (a–c) Longyearbyen (LYR) auroral intensity in the line of sight of the three satellites. (d–f) Wavelet power spectra of 50 Hz raw phase from the Longyearbyen receiver on a decibel (logarithmic) color scale. (g–i) Ny-Ålesund (NYA) line-of-sight auroral intensity. (j–l) Wavelet power spectra from the Ny-Ålesund receiver. (m–o) 1 s $S_4$ amplitude scintillation index from both receivers. (p–r) 1 s $\sigma_\phi$ phase scintillation index from both receivers. (s–u) 60 s VTEC from the satellites at both receivers. (v) GNSS satellites and a selected 630.0 nm all-sky image on a polar axis (zenith angle versus azimuth; magnetic north is up, magnetic east is right).

events (22 December 2014 22:10, 15 January 2015 21:30, 16 January 2015 22:30, and 16 January 2015 23:40). These events were smaller and more transient and complex, which is why we chose to focus on a single, clear event in this study. It bears mentioning, however, that the other events seemed to support the above findings, and that we found no results in contradiction with those of the current study.

## 4. Discussion
### 4.1. Evidence of Irregularities

There is evidence of irregularities in relation to the arc. The presence of SuperDARN backscatter and the enhanced backscatter power indicates the presence of decameter field-aligned irregularities near the arc. There is also enhanced spectral width in relation to the arc. The physical reasons behind enhanced spectral width are complex, but the enhancement might suggest ongoing irregularity processes [*André et al.*, 2000].

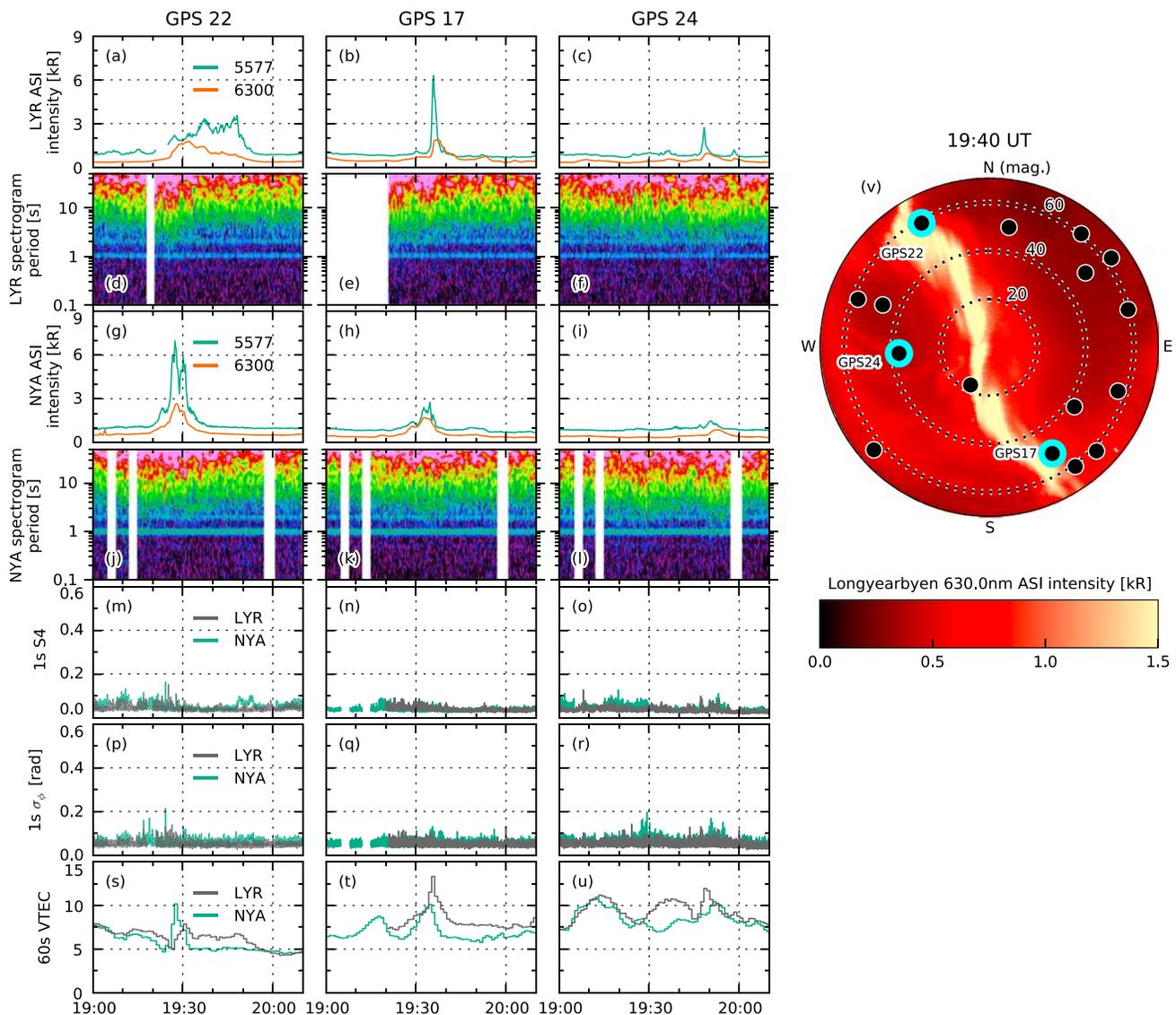

**Figure 6.** See Figure 5 for description. This figure highlights three different satellites.

Another evidence of increased structuring in relation to the arc is the slight enhancements in scintillation and the increase in spectral power in the phase spectrograms. Although the scintillation levels are very low, there are slight increases in phase scintillation in several satellites when the arc crosses the line of sight (as seen in Figures 5 and 6). This indicates that there are indeed some enhancement of irregularities in relation to the arc. This is naturally also visible in the phase spectrograms (since they are derived from the exact same data as the $\sigma_\phi$ index), which show slightly enhanced phase variations down to 1 s. *van der Meeren et al.* [2014] used such spectrograms to study a drifting plasma structure which could be assumed to be stable (not changing in time) compared to its drift speed. It was, therefore, possible to convert the temporal scale of the spectrum to spatial scale by using the drift speed of the plasma. In the current study (as in *van der Meeren et al.* [2015]) we are looking at ionization from particle precipitation, with a significant contribution from the E region (as evidenced by the much stronger 557.7 nm emissions occurring at lower altitudes than the weaker 630.0 nm emissions). The recombination rate at E region altitudes is much higher than in the F region, and the ionized plasma cannot be assumed to be a stable structure drifting westward with the arc. Therefore, we cannot convert the temporal spectrogram scale to spatial scale in this case.

In general, the observations indicate the presence of decameter-scale irregularities in relation to the polar cap arc. From the power law spectrum of irregularities in the ionosphere we can then infer that irregularities must exist at larger scale lengths, too. The irregularities are, however, too weak to cause scintillation effects of concern to GPS users. The presence of strong HF backscatter colocated with low levels of amplitude scintillation has been noted in previous studies [*van der Meeren et al.*, 2014, 2015].

### 4.2. Irregularity Mechanisms

An extensive discussion of all possible structuring mechanisms is outside the scope of this study. However, we would like to briefly discuss some important candidates.

Polar cap arcs are known to be associated with flow shears [e.g., *Carlson and Cowley*, 2005; *Eriksson et al.*, 2006]. The velocity shear seen in the SuperDARN data (Figures 4i and 4j) could drive the Kelvin-Helmholtz (KH) instability. The KH growth rate can be estimated by $\gamma_{KH} = 0.2\Delta V/L$ where $\Delta V$ is the velocity difference and $L$ is the scale length of the velocity difference [*Carlson et al.*, 2008]. Based on the velocity data shown in Figure 4j, the KH growth time can be estimated to ~70 min, which indicates the KH instability is not effective. However, due to the one-dimensionality and low spatial resolution of the SuperDARN data, this estimate may be wildly off in either direction. Based on the available data for our event, we are unable to provide an accurate estimate of the KH growth time.

The gradient drift instability is another common instability mechanism in the polar ionosphere, which works on the trailing edge of drifting plasma structures [e.g., *Tsunoda*, 1988]. It was pointed out in the previous section that there is no single plasma structure drifting westward with the arc, but rather a continuous ionization and recombination which moves westward. Therefore, we assume there will be no structuring by the gradient drift instability.

Finally, we would like to mention the creation of irregularities directly by structured ionization. Previous studies have observed that soft electron precipitation may create $F$ region irregularities [*Kelley et al.*, 1982; *Moen et al.*, 2002]. A statistical study by *Kinrade et al.* [2013] found a relationship between GPS phase scintillation and optical auroral emissions at 557.7 nm ($E$ region). Auroral precipitation is known to be structured on spatial scales down to tens of meters [e.g., *Sandahl et al.*, 2008, and references therein], so it is conceivable that the irregularities are directly created by the auroral precipitation associated with the arc. Unfortunately, we have no way of investigating this further based on the available data from this event.

### 4.3. Electron Density Considerations

A possible explanation for the lack of strong irregularities might be the low density of the plasma. The VTEC measurements (Figures 5s–5u and 6s–6u) indicate a background density of ~5 TECU. This is supported by ionosonde data from Longyearbyen as well as global TEC data from the Madrigal database (not shown). The low density naturally limits the severity of irregularities formed by instability processes. This is in contrast to southward IMF conditions, where dayside density plasma is prominent in the polar cap and the instability processes often develop strong irregularities causing severe scintillation. It is possible that the dayside (and thus higher density) nature of the observations by *Prikryl et al.* [2015a] has contributed to the scintillation levels they observed. Studies of the equatorial ionosphere have shown that the background density can influence the severity of scintillation [*Whalen*, 2009; *Li et al.*, 2011].

Another salient point concerning the electron density is the fact that the arc shows an increase in up to 5 TECU. The total electron content thus reaches twice the background level, and ionosonde data from Longyearbyen show a similar increase (factor of 2) in the $F_2$ peak critical frequency. This meets the relative density requirements of polar cap patches (though the arc is of course not a patch), which are associated with strong scintillation [e.g., *Carlson*, 2012, and references therein]. A review by *Carlson* [1994] points out that scintillation intensity scales with absolute fluctuation in TEC, not relative fluctuation. Thus, even though the arc displays a factor of 2 increase in TEC, the increase is still smaller in absolute terms than high-density patches with dayside densities. It is possible that this is part of the reason why there is no strong scintillation in relation to the arc.

Previous studies have shown that during transitions from IMF south to IMF north, polar cap arcs and patches can coexist due to the sudden appearance of polar cap arcs during IMF north conditions compared to the slower response of patches exiting the polar cap [*Valladares et al.*, 1998; *Basu and Valladares*, 1999]. This could mean that during such transition states, flow shears associated with polar cap arcs may structure the existing high-density plasma (such as patches) through the KH instability. It is thus conceivable that polar cap arcs in a high-density ionosphere could pose a problem with regards to scintillation. In the current study the IMF was consistently northward oriented several hours before the event, so the transition effect is not relevant (as confirmed by the background density measurements). It would be beneficial for future studies to study plasma structuring associated with polar cap arcs when the background density is high, such as in summer or during transition states from IMF north to IMF south.

## 5. Conclusions

This study has shown a single case of scintillation and irregularities in relation to a polar cap arc in unprecedented detail. The main findings of the study can be summarized as follows.

1. During northward IMF conditions, weak irregularities are observed in relation to the nightside part of a long-lived, Sun-aligned polar cap arc.
2. The irregularities are evidenced by the presence of HF backscatter (indicating decameter-scale irregularities) as well as slight enhancements of GNSS phase variations and 1 s $\sigma_\phi$.
3. The polar cap arc causes no amplitude scintillation and does not cause significant phase scintillation ($\sigma_\phi \leq 0.2$).
4. The lack of scintillation may be related to the low density in the polar cap after several hours of northward IMF.

Further studies should be carried out to close the gap in our knowledge of high-latitude irregularities during northward IMF conditions. Specifically,

1. further case studies and statistical studies should be carried out to get a representative view of irregularities and scintillation in relation to polar cap arcs;
2. direct comparisons should be made between auroral arcs and polar cap arcs of similar intensities, to find out if there is a difference between the effects of irregularities created by precipitation in the auroral oval and in the polar cap; and
3. polar cap arcs should be studied when there are high background densities, which could enable more severe irregularity formation and possibly a stronger impact on GNSS scintillation. This could be done by studying polar cap arcs after a south-to-north transition of the IMF, where the polar cap may be filled with high-density dayside plasma when the arc forms. Alternatively, one could study the dayside portion of a polar cap arc or arcs during summer.

**Acknowledgments**
The University of Oslo ASI data are available at http://tid.uio.no/plasma/aurora. The IMF and solar wind data were provided by the NASA OMNIWeb service (http://omniweb.gsfc.nasa.gov). The DMSP SSUSI data are available from http://ssusi.jhuapl.edu. The DMSP particle detectors were designed by Dave Hardy of AFRL and data obtained from JHU/APL. The SuperDARN data are retrieved from Virginia Tech using the DaViTpy software package. The ionosonde data (not shown) are available from http://dynserv.eiscat.uit.no. The GNSS data can be made available upon request from the author. This study was supported by the Research Council of Norway under contracts 212014, 223252, and 230935. We wish to thank Nikolai Østgaard at the Birkeland Centre for Space Science, University of Bergen for helpful discussions.